\documentclass[11pt]{article}

\usepackage{a4wide}
\usepackage{wrapfig}
\usepackage{floatflt}
\usepackage{amssymb}
\usepackage{amsmath}
\usepackage{graphicx}
\usepackage{xspace}
\usepackage{url}

\newcommand{\as}{\alpha_s}

\newcommand{\order}[1]{{\mathcal O}\left(#1\right)}
\newcommand{\ie}{i.e.\ }
\newcommand{\eg}{e.g.\ }

\newcommand{\PU}{\mathrm{PU}}

\newcommand{\ns}{\,\text{ns}}

\newcommand{\cm}{\,\text{cm}}
\newcommand{\s}{\,\text{s}}
\newcommand{\GeV}{\,\text{GeV}}
\newcommand{\TeV}{\,\text{TeV}}
\newcommand{\pbpb}{{Pb\,Pb}\xspace}
\newcommand{\Erf}{\mathrm{Erf}}

\title{
  \textbf{Pileup subtraction using jet areas}
}

\author{
  Matteo~Cacciari and Gavin~P.~Salam\\
  \small \it LPTHE, Universit\'e P. et M. Curie -- Paris 6,\\
  \it\small 
  Universit\'e D. Diderot -- Paris 7, CNRS UMR 7589,
  Paris, France
}

\date{}

\begin{document}


\maketitle

\vspace{-6cm}
\begin{flushright}
  July 2007\\
  LPTHE-07-01\\
\end{flushright}
\vspace{4.5cm}
 
\noindent \textbf{Abstract: }{%
  One of the major challenges for the LHC will be to extract precise
  information from hadronic final states in the presence of the large
  number of additional soft $pp$ collisions, pileup, that occur
  simultaneously with any hard interaction in high luminosity runs. We
  propose a novel technique, based on jet areas, that 
  provides jet-by-jet corrections for pileup and underlying-event
  effects. It is data driven, does not depend on Monte Carlo modelling
  and can be used with any jet algorithm for which a jet area can be
  sensibly defined.  We illustrate its effectiveness for some key
  processes and find that it can be applied also in the context of the
  Tevatron, low-luminosity LHC and LHC heavy-ion collisions.%
} \bigskip

\section{Introduction}
\label{sec:intro}

The Large Hadron Collider (LHC) will collide protons with an
unprecedented instantaneous luminosity of up to
$10^{34}\cm^{-2}\s^{-1}$ and a bunch 
spacing of $25\ns$.  While this high luminosity is essential for many
searches of rare new physics processes at high energy scales, it also
complicates 
analyses, because at each bunch crossing there will be of the order of 20
minimum bias $pp$ interactions, which pollute any interesting hard
events with many soft particles. The beams at LHC will have a
longitudinal spread, and it may be possible experimentally to associate
each charged particle with a distinct primary vertex that corresponds
to a single $pp$ interaction and so eliminate some fraction of the
soft contamination. However, for neutral particles this is not
possible, and most jet measurements are in any case expected to be
carried out with calorimeters, which do not have the angular
resolution needed to reconstruct the original primary vertex.
Therefore kinematic measurements for jets will be adversely affected
by pileup (PU), with resolution and absolute energy measurements
suffering significantly.

Both the Tevatron and LHC experiments have examined the question of
pileup. Some approaches are based on average correction procedures,
for example the requirement that final measured distributions should
be independent of luminosity~\cite{CDFkt}, or a correction to each jet
given by some constant times the number of primary interaction
vertices (minus one)~\cite{CDFCone}. These approaches have the
advantage of being simple, but their averaged nature limits the extent
to which they can restore resolution lost through pileup.
Other approaches involve event-dependent corrections that are applied
to calorimeter towers either before or during the
clustering~\cite{CMS-sub,AliceHija}. While these can give better
restoration of resolution than average-based methods, they are tightly
linked to the specific experimental setup (for example calorimeter
cell-size), and require ad-hoc transverse-momentum thresholds to
distinguish pileup from hard jets. Additionally they are sometimes tied to
specific (legacy) jet algorithms, and may not always be readily
applied to the more modern jet algorithms that are increasingly being
adopted in the hadron-collider communities.

\section{The method}
\label{sec:method}

We propose an event-by-event, and jet-by-jet, pileup subtraction
approach that is suitable for any infrared safe jet algorithm, and
which performs the corrections {\it after} the jet finding has been
carried out, so as to ensure independence on the specific detector
setup. It is based on two novel ingredients: {\it i}) the measurement
of each jet's susceptibility to contamination from diffuse noise and
{\it ii}) an essentially parameter-free technique to measure the
level, $\rho$, of this diffuse noise in any given event, where by
noise (or also `background'), we refer to any form of diffuse
contamination in the event, usually due to minimum-bias pileup and to
some extent 
the underlying event.\footnote{The terms noise and background are
  specifically not intended to refer to experimental (e.g.\
  electronics) noise, though it is not inconceivable that such
  experimental noise could be treated on a similar footing.}

{\it i}) The jet's susceptibility to contamination is embodied in the
jet \emph{area}, $A$, measured on the rapidity ($y$), azimuth ($\phi$)
cylinder. The jet area is a non-trivial (and novel) concept insofar as
a jet consists of pointlike particles, which themselves have no
intrinsic area.  To define a sensible area one therefore adds
additional, infinitely soft particles (ghosts) and identifies the region in
$y$, $\phi$ where those ghosts are clustered with a given jet.
The extent of this region gives a measure of the (dimensionless) jet
area.

The jet area is different for each jet and depends on the details of
its substructure, and to some extent on the event as a whole. Contrary
to common wisdom, jet areas can differ significantly from $\pi R^2$
even for more geometrical jet definitions, such as cone algorithms
($R$ here is the radius parameter in the jet definition).
Consequently only the measured area for each jet provides reliable
information about its level of potential contamination.

The exact details of the determination of the jet areas (for example the
choice of distribution of infinitely soft particles) are largely
irrelevant here, and are instead to be found discussed at length
in~\cite{css-area}, which also presents studies of the properties of
jet areas for a range of jet algorithms.
The practical measurement of the jet areas is carried out using the
{\tt FastJet} package~\cite{FastJetProg} and relies on the fast
computational strategies for jet clustering described
in~\cite{fastjet,siscone}.

Given a suitable definition of jet area,
the modification of a jet's transverse momentum ($p_t$) by diffuse
noise can be shown to be~\cite{css-area}
\begin{equation}
  \label{eq:deltapt}
  \Delta p_t =  A \rho \pm \sigma \sqrt{A} - L\,,\qquad \langle
  L\rangle = \order{\as \cdot A \rho \,\ln \frac{p_t}{A \rho} },
\end{equation}
where $\rho$, the level of diffuse noise, corresponds to the amount of
transverse momentum added to the event per unit area, for example by
minimum 
bias particles. These particles are taken to be dense on the scale $R$
of the jet algorithm, as is bound to be the case with many minimum
bias events, and $\sigma$  is the
standard deviation of the resulting noise when measured across many
regions of unit area.
At high-luminosity at LHC $\rho$ is expected~\cite{Pythia} to be
$\sim 10-20\GeV$ per unit area. %
The first term in eq.~(\ref{eq:deltapt}) is therefore the geometrical
contamination of the jet and is associated with an uncertainty (second
term) because of fluctuations in the noise from point to point in the
event.
The third term, $L$, accounts for the occasional loss (or the even
more occasional gain) of part of the jet's contents, due to the fact
that jets can be modified when clustered in the presence of diffuse
noise, as some of the particles originally clustered into one jet can
instead end up in a different one.
One should be aware that this contribution has a very non-Gaussian
distribution --- usually it is small, but a fraction $\as$ of the time
it can become comparable to $A \rho$.%
\footnote{As discussed
  in~\cite{css-area}, the average value of $L$ is dominated by
  situations in which an emission $p_2$ is near the edge of the jet
  with $A \rho \ll p_{2t} \ll p_t$ and is lost from the jet. This is a
  very rare occurrence, $\sim \as dp_{2t} /p_{2t} \cdot
  (A\rho/p_{2t})$, with the second (suppression) factor embodying the
  fact that as the emission $p_2$ is made harder it is progressively
  more difficult for its clustering fate to be altered by the minimum
  bias momenta. After integration over $p_{2t}$ these very rare
  occurrences give the dominant (logarithmic) contribution to $\langle
  L \rangle$ because of the weighting with the resulting change in jet
  momentum, $p_{2t}$. Numerical investigations indicate that for both
  the $k_t$ and Cambridge/Aachen algorithms $\langle L \rangle \simeq
  \frac{0.3 \as C}{\pi} A \rho\, \ln \frac{p_t}{A \rho}$, where $C$ is
  $C_F$ or $C_A$ according to whether the jet is quark-like or
  gluon-like, implying rather small average effects.}

Our correction procedure will be based on the assumption that
fluctuations are small ($\sigma \ll \sqrt{A} \rho$) and on the idea that one
can neglect the loss term $L$ of eq.~(\ref{eq:deltapt}), on the grounds
that for the majority of events it is much smaller than $A\rho$.
We will therefore correct the measured $p_t$ of each jet $j$ via the
subtraction:
\begin{equation}
  \label{eq:pt-correct}
  \smash{p_{tj}^{(\text{sub})}} = p_{tj} - A_j \rho \,,
\end{equation}
where $A_j$ is that jet's area.

Eq.~(\ref{eq:pt-correct}) provides a correction to the jet's scalar
transverse momentum. There can be situations in which an observable is
sensitive to the jet's direction, and more generally where one needs
to correct the jet's full 4-vector (for example for large jet radii,
where the contamination from the background can build up a significant
invariant mass). In these cases eq.~(\ref{eq:pt-correct}) may be
extended to full 4-vector form, making use of a `4-vector' area
$A_{\mu j}$: 
\begin{equation}
  \label{eq:pt-correct-4vect}
  \smash{p_{\mu j}^{(\text{sub})}} = p_{\mu j} - A_{\mu j}\, \rho \,.
\end{equation}
The precise definition of the 4-vector area is provided
in~\cite{css-area}, but essentially it can be understood as the
integral of a massless 4-vector over the surface of the jet,
normalised such that its transverse component $A_{t j}$ coincides with
the scalar area $A_{j}$ for small jets. It is this 4-vector correction
(supplemented with a `sanity check' which for example removes all jets
for which $p_{tj} \le A_{tj} \rho$) %
that has been used in figs.~\ref{fig:Zpmass}, \ref{fig:Wttbar}
and~\ref{fig:Tevatron} below. We note that the $\rho$ used for the
$4$-vector correction is the same (scalar) quantity as used in
eq.~(\ref{eq:pt-correct}).

A point to be borne in mind is that the procedure used here for
defining areas can only be applied to all-order infrared-safe jet
algorithms.  This is because the jet area is meaningful only if the
hard-particle content of the jet is unaltered by the addition of the
soft ghost particles. Certain jet algorithms are not infrared safe,
notably seed-based iterative and midpoint cone algorithms with a zero
seed-threshold, and therefore cannot be used in this context. Areas
can in contrast be defined for collinear unsafe algorithms (e.g.
algorithms with a finite seed threshold). However both infrared and
collinear unsafe algorithms are in any case highly deprecated because
of the divergences that appear for them in perturbative calculations,
and because of their related instability with respect to
non-perturbative effects.

{\it ii}) The second ingredient in carrying out the subtraction is the
estimate of $\rho$ for each event. The principal difficulty in
estimating the amount of noise is that of distinguishing the noise
component from the large amounts of $p_t$ deposited by the hard event;
for example one cannot simply take the ratio of the total amount of
$p_t$ in the event divided by the full area over which one measures.
It turns out however that some jet algorithms, like
$k_t$~\cite{kt} and Cambridge/Aachen~\cite{cam} (but not
SISCone~\cite{siscone}), lead to a large sample
of quite regular soft pileup `jets' for each event --- these jets do
not represent any particular hard structure in the pileup, but rather
reflect these jet algorithms' tendency to naturally organise a uniform
background of soft particles into structures (`jets') each of area $\sim \pi
R^2$.
In the limit in which the noise component is uniform and dense,
each pure pileup jet will have the property that its $p_t$ divided by its
area is equal to $\rho$. In practice pileup has local
fluctuations, causing the $p_{tj}/A_j$ values to be
distributed around $\rho$. We propose that one measure $\rho$ for each event
by taking the median value of this distribution in the event:%
\begin{equation}
  \label{eq:rho}
  \rho = \text{median}\left[ \left\{ \frac{p_{tj}}{A_{j}}\right\}
  \right].
\end{equation}
One may in an analogous way extract $\sigma$, giving it a value such
that a fraction $(1-X)/2$ of jets have $p_{tj}/A_j < \rho -
\sigma/\sqrt{A_j},$ where $X=\Erf(1/\sqrt{2}) \simeq0.6827$ is the
fraction of a Gaussian distribution within one standard deviation of
the mean. We use a one-sided determination of $\sigma$ (rather than a
symmetric requirement that a fraction $X$ of jets satisfy
$\rho-\sigma/\sqrt{A_j} < p_{tj}/A_j < \rho+\sigma/\sqrt{A_j}$),
because it is expected to be less sensitive to bias from hard jets.
Note also that for simplicity, in practice we replace $\sqrt{A_j}$
with $\sqrt{\langle A\rangle}$

The above pileup subtraction procedure is valid as long as three
conditions hold:
\begin{enumerate}
\item  The pileup noise should be independent of rapidity and
  azimuth. If it isn't the procedure can be extended appropriately.
\item The radius $R$ of the jet algorithm should be no smaller than
  the typical distance between minimum bias particles, otherwise the
  extraction of $\rho$ from the median will be biased by the large
  amount of empty area not associated with jets. This condition may be
  relaxed if one also allows the ghosts used for area measurements to
  cluster among themselves to form `pure ghost jets' (jets free of any
  hard particles)~\cite{css-area}.  These pure ghosts jets (which have
  infinitesimal transverse momentum) can then be included in the set
  of jets used to calculate the median in eq.~(\ref{eq:rho}),
  providing a way to account for significant area that is free of hard
  particles. For $\rho$ to be a reasonably reliable estimate of the
  noise density one should then require that $\sigma$ be smaller than
  $\rho$.

  \item The number of pileup jets should be much larger than
  the number of jets from the hard interaction that are above scale
  $\langle A\rangle \rho$,
  \begin{equation}
    \label{eq:median-works-simple}
     n_\PU \,\gg\, n_{H}^{p_t > \langle A\rangle \rho}\,.
  \end{equation}
  If this is not the case, then the median will be significantly
  biased by the hard jets. 
  In a first-order, fixed-coupling, leading logarithmic approximation
  for a central dijet event, considering just independent emission
  from the incoming partons, we have
  \begin{equation}
    \label{eq:nH}
    n_{H}^{p_t > \langle A\rangle \rho}
    \simeq 2 + \frac{2\as C}{\pi}
    \int_{-y_{\max}}^{y_{\max}} dy \int_{\langle A\rangle \rho}^{Q
      e^{-y}} \frac{d p_t}{p_t}\,,
  \end{equation}
  where the colour factor $C$ is $C_A=3$ or $C_F=4/3$, according to
  the nature of the incoming partons and $y_{\max}$ is the maximum
  rapidity being considered (we assume $y_{\max} < \ln \frac{Q}{\langle
    A\rangle\rho}$). Using $n_\PU \simeq 4\pi y_{\max}/\langle
  A\rangle$, we then obtain the requirement,
  \begin{equation}
    \label{eq:median-works}
    \frac{\as C}{\pi^2}  \left(\ln
      \left(\frac{Q}{\langle A\rangle\rho}\right) -
      \frac{y_{\max}}{2}\right) \cdot\,\langle A\rangle 
    \;\sim \;
    \frac{\as C}{\pi}  \ln \left(
      \frac{Q}{\pi R^2 \rho}\right)  \cdot\, R^2
    \; \ll \;
    1 \,,
 \end{equation}
 where the second level of approximation uses $\langle A\rangle \sim
 \pi R^2$.
\end{enumerate}
The second and third conditions above place lower and upper limits on
$R$ that depend on the luminosity (specifically the number density of
particles) and on the hard scale $Q$.  We find, in practice, that at
very low luminosity (a few tens of particles per unit rapidity, $\rho
\simeq 2-5\GeV$) $R$ should be in the range $0.5-0.6$ for the estimate
of $\rho$ to be reliable,
while at high luminosity at the LHC (a few hundred particles per unit
rapidity, $\rho \simeq 10-20 \GeV$) one can roughly use the range $0.2
\lesssim R \lesssim 1$ for $Q\simeq 200\GeV$. Note that these
restrictions apply exclusively to the determination of $\rho$: the
value of $R$ (and even the choice of jet algorithm) used here need not
be the same as that used to carry out the main jet-finding and
subtraction, and later in the article we shall make use of this
freedom.

\begin{figure}[tp]
  \includegraphics[width=0.48\textwidth]{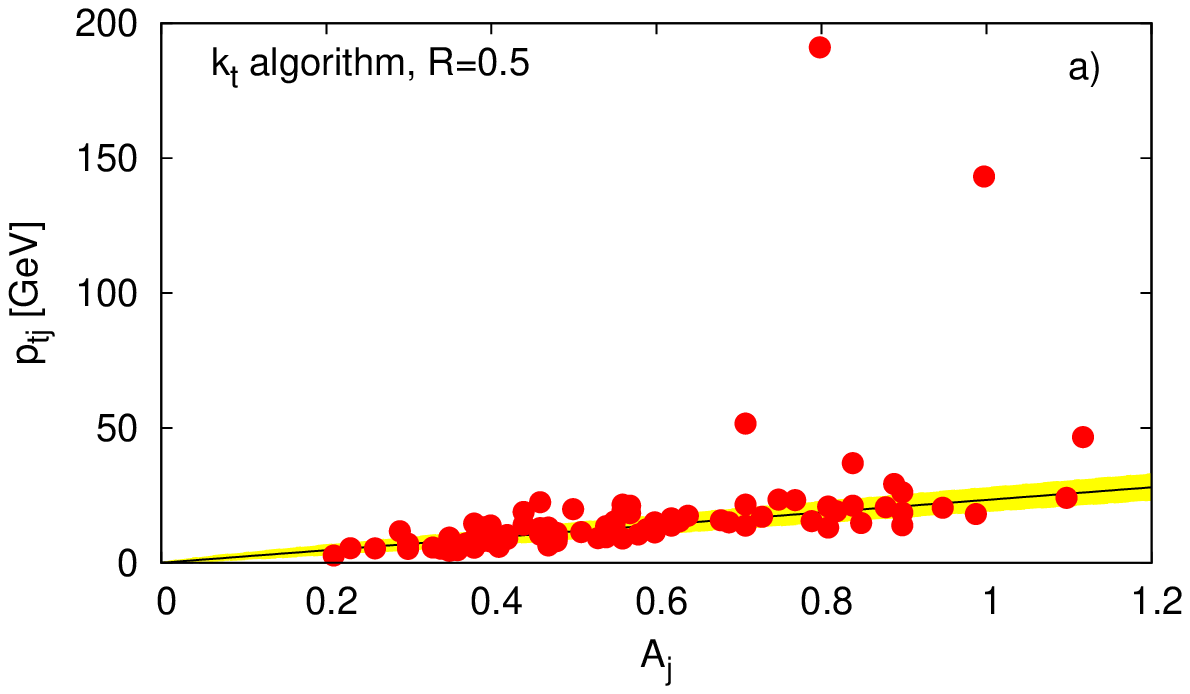}\hfill
  \includegraphics[width=0.48\textwidth]{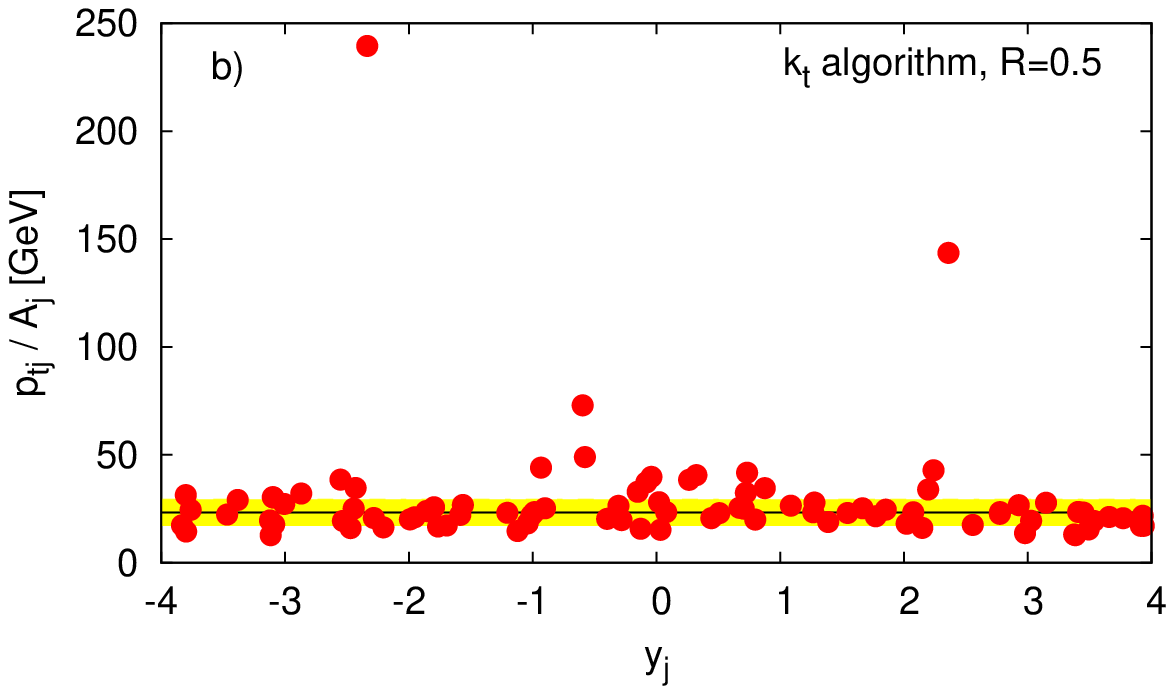}
  \caption{\small a) Scatter plot of the jet transverse momentum
    $p_{tj}$ versus its area $A_j$, for an LHC dijet event with a
    pileup of 22 minimum bias interactions (simulated with the default
    tune of Pythia~6.325~\cite{Pythia}, as is the case for all results
    in this paper, except fig.~\ref{fig:ue-correlation}). The line and
    band are given by $\rho A_j \pm \sigma \sqrt{A_j}$.  b) The ratio
    $p_{tj}/A_{j}$ as a function of the rapidity, $y_j$, for the same
    event; the line and band are given by $\rho \pm \sigma /
    \sqrt{\langle A\rangle}$.}
  \label{fig:pt-v-area}
  \label{fig:pt_over_area-v-rap}
\end{figure}

To help illustrate the extraction of $\rho$,
figure~\ref{fig:pt-v-area}a shows a scatter-plot of $p_t$ versus area
for each jet found in a single high-luminosity LHC event simulated
with Pythia~\cite{Pythia}. It contains one hard dijet event and $22$
minimum-bias interactions. One sees a clear linear correlation between
$p_{tj}$ and $A_j$, except
in the case of a few hard jets.  Figure~\ref{fig:pt_over_area-v-rap}b
shows the ratio of $p_{tj}/A_j$ for each jet as a function of its
rapidity $y_j$ --- with the exception of the few hard jets,
the typical values are clustered around a rapidity-independent value
that coincides closely with $\rho$ as extracted using the median
procedure, eq.~(\ref{eq:rho}).
These features, while illustrated just for the $k_t$
algorithm~\cite{kt} with $R=0.5$,
are found to be similar with other values of $R$ and for the
Cambridge/Aachen algorithm~\cite{cam}.

\begin{figure}[t]
  \includegraphics[width=0.48\textwidth]{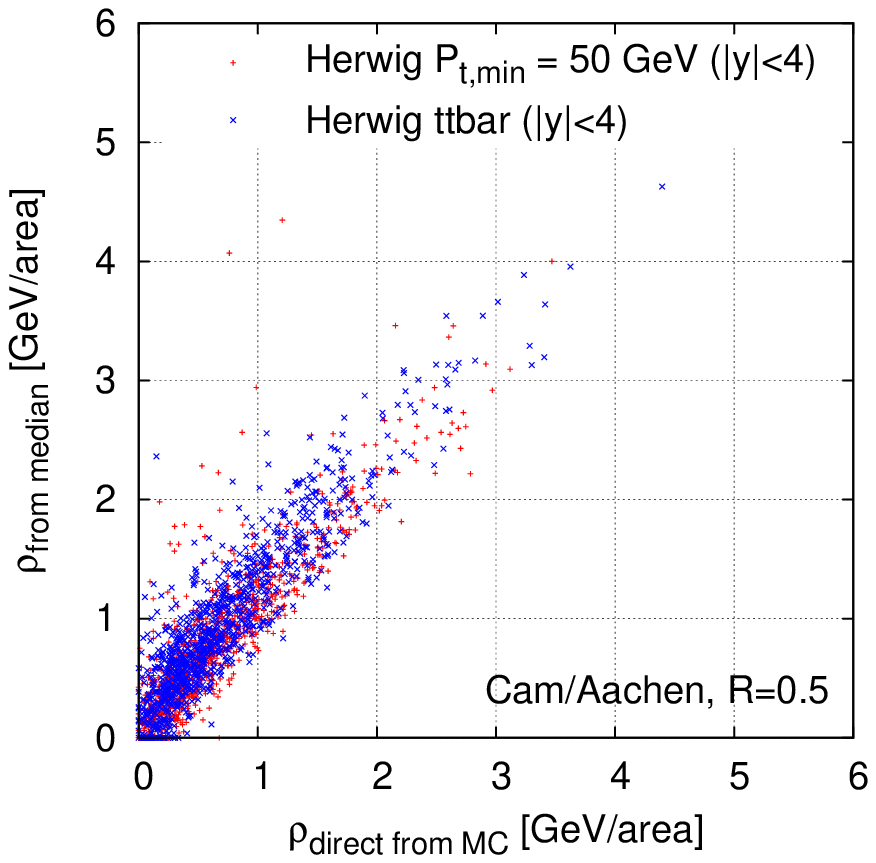}\hfill
  \includegraphics[width=0.48\textwidth]{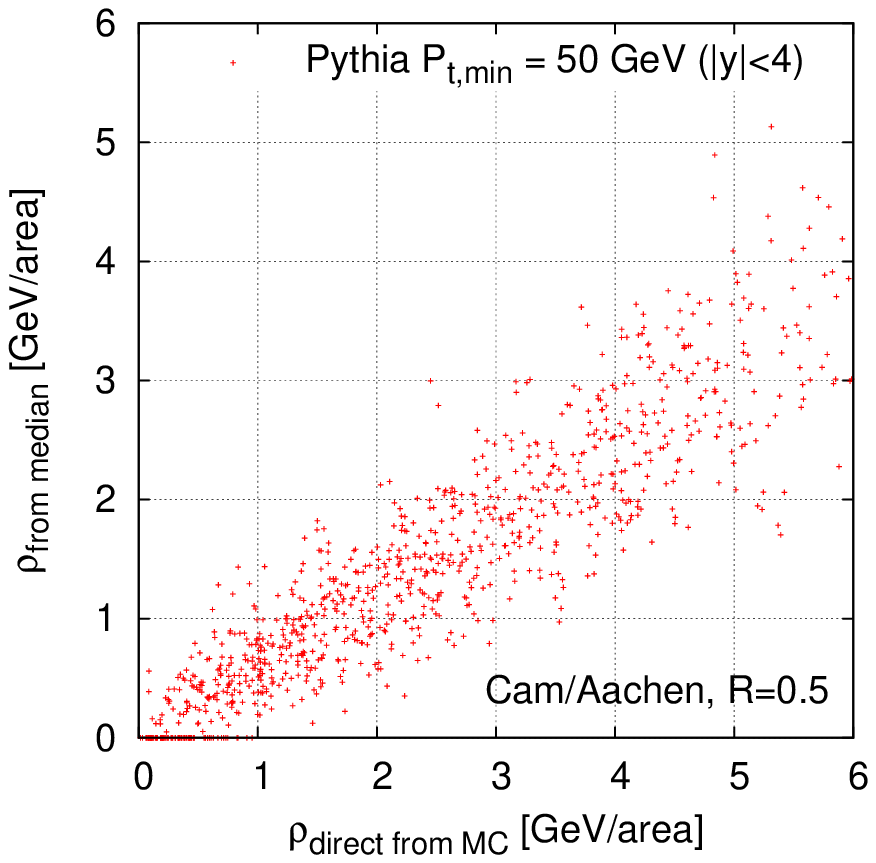}
  \caption{\small a) Scatter plot of the value for $\rho$ extracted by
    using eq.~(\ref{eq:rho}) versus the transverse momentum per unit
    area added by Herwig's UE to dijet production and $t\bar t$
    production at the LHC (version 6.510, default tune).  b) The same
    correlation using Pythia dijet events (version 6.412, default
    tune). In the case of Herwig the UE contribution added is that of
    the hadrons produced in the soft-underlying event stage; in the
    case of Pythia, hadrons cannot be unambiguously ascribed to the
    hard or underlying events so instead we consider the total (scalar
    sum) $p_t$ of the hadrons minus that of the perturbative partons
    (all partons that enter strings, and that are connected via
    quark/gluon lines to the hard scatter).
    Fitting a straight line $\rho_\text{from median} = a+b
    \rho_\text{direct from MC}$ to the data sets yields $a =
    0.13\pm0.02\GeV$, $b = 1.03\pm0.02$ for the Herwig dijet events
    and $a=-0.02\pm 0.02\GeV$ and $b = 0.62\pm 0.01$ for the Pythia
    dijet events. The departure from $b=1$ in the case of Pythia is
    probably due to a sizeable pointlike component (\ie extra jets) in
    Pythia's underlying event, while $\rho$ as measured by our method
    reflects only the part of the underlying event that is diffuse on
    the scale of the jet radius $R$.
  }
  \label{fig:ue-correlation}
\end{figure}

A point worth bearing in mind in the extraction of $\rho$ is that
diffuse radiation comes not only from pileup, but also from the
underlying event (UE), and $\rho$ inevitably receives contributions
from both. Thus any subtraction using $\rho$ will remove both pileup
and the diffuse part of the UE. An interesting corollary to this is
that in events without pileup the method provides a novel way to
directly measure the diffuse component of the underlying event on an
event-by-event basis. Monte Carlo studies show that this measure is indeed
closely correlated with the Monte Carlo input for the UE, as shown in 
fig.~\ref{fig:ue-correlation} for both Herwig and Pythia.
This suggests that the method that we propose in this paper can be
fruitfully employed also to study the underlying event.

This measure
of the diffuse part of the UE is rather special in that purely
perturbative events lead to a non-zero value for $\rho$ only at
extremely high orders: for $\rho$ to be non-zero, there must be at
least as many perturbative jets as pure-ghost jets; assuming that $n$
perturbative particles lead to $n$ perturbative jets (a rare but
possible occurrence), using the result~\cite{css-area} that the
typical areas of single-particle jets and pure ghost jets are roughly
$x_{\text{sp}} \pi R^2$ and $x_{\text{pg}} \pi R^2$ respectively (with constants $x_{\text{sp}}
\simeq 0.81$, $x_{\text{pg}} \simeq 0.55$ for both $k_t$ and
Cambridge/Aachen), and requiring the total area of the 
jets to sum up to $4 \pi y_{\max}$, one obtains the following
approximate result for the minimal perturbative order $n$ for $\rho$
to be non-zero:
\begin{equation}
  \label{eq:rho-non-zero}
  n \;\simeq\; \frac{4 y_{\max}}{(x_{\text{pg}} + x_{\text{sp}})R^2} \;\simeq\; 2.94
  \frac{y_{\max}}{R^2} \,.
\end{equation}
For $y_{\max}=4$ and $R$ in the range $0.5-0.7$, this translates
to the statement that $\rho$ will be zero perturbatively roughly up to
orders $\as^{24}-\as^{47}$. This extends the ideas of
\cite{MarchesiniWebberMinCone} which had pushed the order of
perturbative contamination in the underlying event estimation to
$\as^4$, by considering the activity in the less energetic of two
cones placed in between the hard jets.

\section{Example applications}
\label{sec:appl}

Once one has extracted $\rho$, one can apply
eqs.~(\ref{eq:pt-correct}) or (\ref{eq:pt-correct-4vect}) so as to
correct the momentum of each individual jet. We shall first show three
examples of this in high-luminosity LHC situations: jet transverse
momenta in dijet events, reconstruction of a hypothetical leptophobic
$Z'$, and top mass reconstruction.  Then we will examine results
from a low-pileup example (Tevatron) and a very high-background level
example (heavy-ion collisions).

\begin{figure}[t]
  \includegraphics[width=\textwidth]{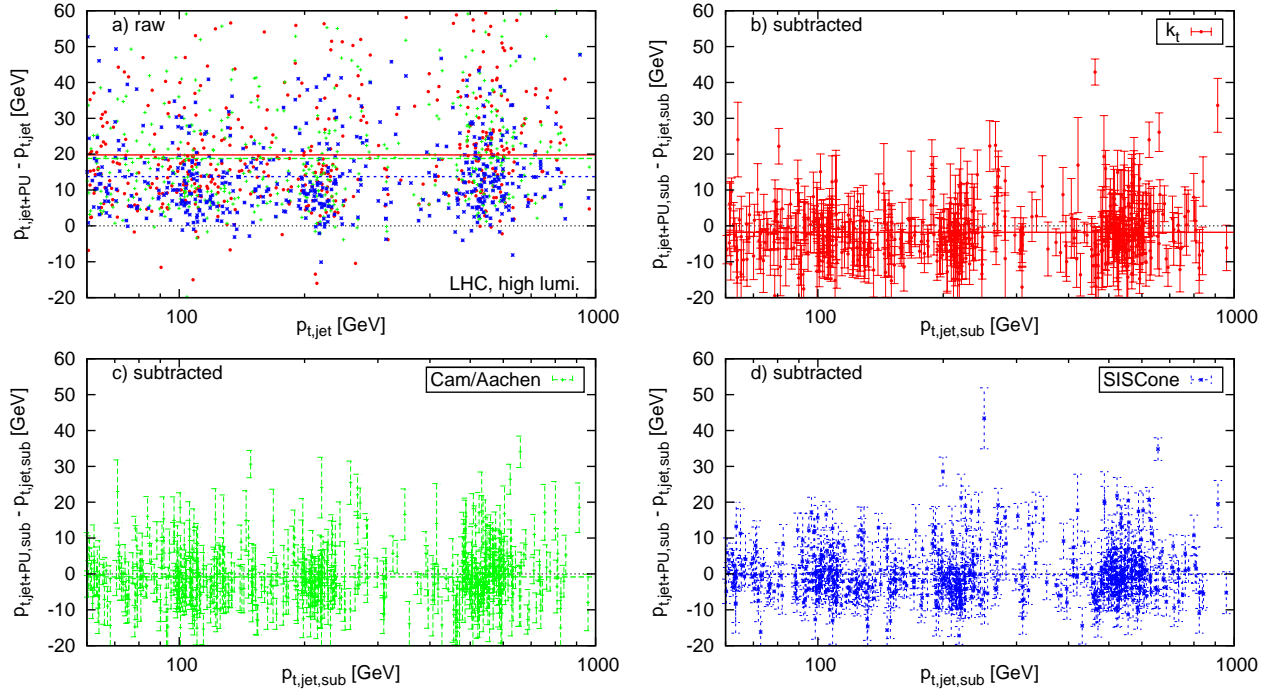}
  \caption{\small Effect of pileup and subtraction on a sample of simulated
    dijet events at LHC, generated with Pythia: a) the points show the
    difference in $p_t$ between jets in the event with and without
    pileup (as a function of jet $p_t$), for three different jet
    algorithms, while the lines correspond to
    a fit of the shifts for each algorithm; b--d)
    similarly for each algorithm individually, but with all jet $p_t$'s
    having been corrected using the 
    subtraction method described in the text, with error bars
    corresponding to the $\sqrt{A} \sigma$ term of
    eq.~(\ref{eq:deltapt}). Subtraction has been applied also to the
    events without pileup, so that the underlying event contribution
    is removed in both cases. The pileup used corresponds to
    high-luminosity running, ${\cal L} =
    10^{34}\,\text{cm}^{-2}\,\text{s}^{-1}$ and a bunch spacing of
    $25\,\text{\ns}$. For all jet algorithms we use $R=0.7$ (and additionally
    $f=0.5$ for SISCone~\cite{siscone});  $\rho$ was obtained using
    the $k_t$ algorithm
    with $R=0.5$. %
    (The observed clustering in certain $p_t$ ranges is a consequence
    of the use of Monte Carlo samples generated with minimal $p_t$
    values of $50$, $100$, $200$ and $500\GeV$.)}
  \label{fig:sub-correl-2tile}
\end{figure}

In fig.~\ref{fig:sub-correl-2tile}a we have taken samples of
simulated dijet events at various transverse momenta and clustered
them both on their own and together with high-luminosity pileup.
To simplify the task of matching the jets clustered with and without
pileup, we reject events (about $\sim 25\%$) in which, without pileup,
a third jet is present and has a transverse momentum greater than half
that of the second hardest jet.
For each selected event, we have identified the two hardest jets, and
plotted the 
shift in each jet's $p_t$ due to the pileup (being careful to properly
match each of the two jets with pileup to the corresponding one
without pileup). The shift is significant (up to $\sim 20\GeV$ on
average) and varies considerably from jet to jet (up to $\sim
50\GeV$), both because of the variation in jet areas and because the
pileup fluctuates from event to event. The negative shifts observed
for a small subset of jets are attributable to the pileup having
modified the clustering sequence, for example breaking one hard jet
into two softer subjets (this is related to the 3rd term of
eq.~(\ref{eq:deltapt})).

Figs.~\ref{fig:sub-correl-2tile}b--d show what happens once we use the
subtraction procedure, eq.~(\ref{eq:pt-correct}). It has
been applied to the events both with and without pileup: even without
pileup there is a non-negligible amount of diffuse radiation, which
comes from the underlying event (UE, $\langle\rho_{UE}\rangle \sim
2.5\GeV$, to be compared with that from only the pileup,
$\langle\rho_{PU}\rangle \sim 14\GeV$) --- the subtraction in the case
with pileup removes both the PU and the UE contributions (the measured
$\langle \rho\rangle \sim 17 \GeV$ cannot distinguish between them),
and it would be inconsistent to compare with jets that still contain
the UE. From the plots, one sees that the average subtracted shift is
now always within $\sim 1\GeV$ of zero. 
The non-uniformity of the pileup causes the jet-by-jet shift to still
fluctuate noticeably (and it is often negative), however the points
are nearly all consistent with zero to within their error bars
($\sqrt{A}\sigma$ in eq.~(\ref{eq:deltapt})). One notes that these
fluctuations are almost a factor of $2$ smaller ($1.5$ for SISCone)
than those before subtraction in fig.~\ref{fig:sub-correl-2tile}a. This
is especially visible for the $k_t$ algorithm whose larger area
variability led to the greatest fluctuations in the unsubtracted case.
The reduction in fluctuations is one of the key strengths of this
approach and would not be obtained in an average-based subtraction
procedure.

\begin{figure}[t]
  \centering
  \includegraphics[width=0.5\textwidth]{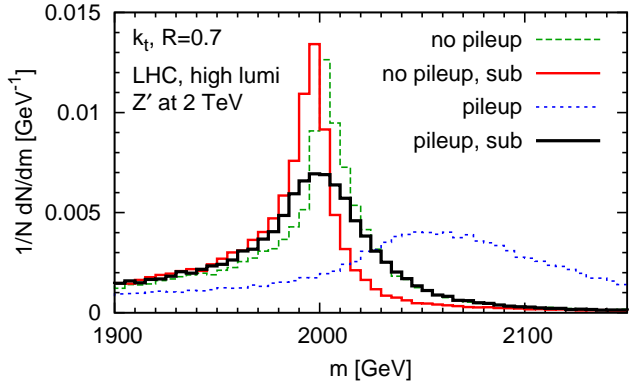}
  \caption{\small Invariant mass distribution of the two hardest jets
    in hadronically decaying $Z'$ events at the LHC, as simulated with
    Pythia~6.325. It illustrates the effect of the subtraction in
    samples with and without high-luminosity pileup ($\rho$ extracted
    using the $k_t$ algorithm with $R=0.5$). Further details in
    text.
  }
  \label{fig:Zpmass}
\end{figure}

\begin{table}[tbp]
  \centering
  \begin{tabular}{l|cc|cc|cc|}
                   & \multicolumn{2}{c|}{$k_t$} 
                   & \multicolumn{2}{c|}{Cam/Aachen} 
                   & \multicolumn{2}{c|}{SISCone} \\
                   & $m$ & $\Delta m$ 
                   & $m$ & $\Delta m$ 
                   & $m$ & $\Delta m$ \\\hline
    no pileup      & 2003 & 10 & 2002 & 10 & 1998 & 10 \\
    no pileup, sub & 1995 & 13 & 1995 &  8 & 1993 & 10 \\
    pileup         & 2065 & 60 & 2049 & 48 & 2030 & 33 \\
    pileup, sub    & 1998 & 25 & 1998 & 25 & 1997 & 20 \\
  \end{tabular}
  \caption{Reconstructed masses and widths (in GeV) for the $Z'$ peak
    (cf.\ fig.~\ref{fig:Zpmass}) with and
    without pileup, and with and without subtraction; $\Delta m$ is
    the half-width at half peak-height, while $m$ is the average mass
    determined in the part of the distribution within half
    peak-height. The accuracy of the results is $\sim 3\GeV$, limited
    mainly by binning artefacts.}
  \label{tab:Zp}
\end{table}

\begin{figure}[t]
  \centering
  \includegraphics[width=0.48\textwidth,height=10.5cm]{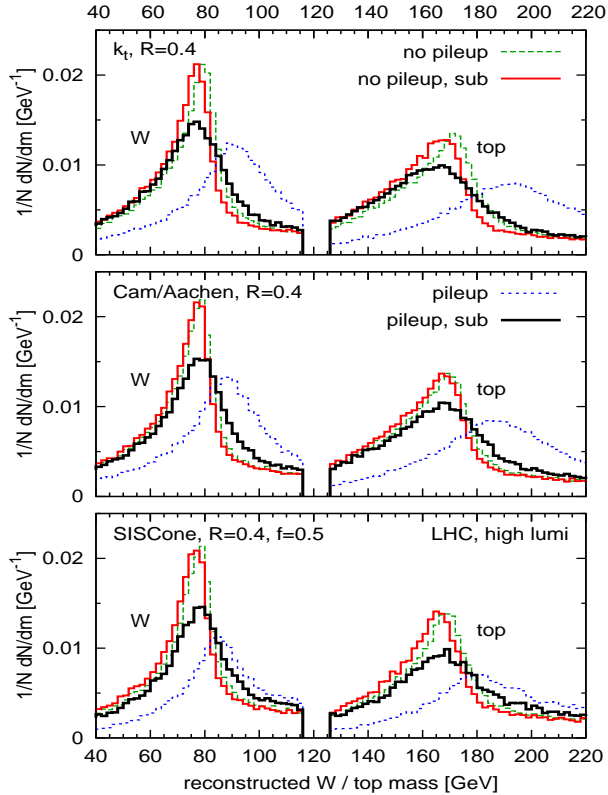}\hfill
  \caption{\small $W$ and top mass reconstruction in simulated LHC
    $t\bar t$ events (decaying to a lepton plus jets), illustrated
    with and and without pileup.  Simulated with Pythia~6.325, with
    $m_\mathrm{top} = 175 \GeV$. 
    We have extracted $\rho$ using the same jet definition as for the
    $W/t$ reconstruction, except in the case of SISCone for which
    $\rho$ was extracted with the $k_t$ algorithm and $R=0.5$. %
  }
  \label{fig:Wttbar}
\end{figure}

Our next high-luminosity LHC study is more physical and considers the
reconstruction of a hypothetical leptophobic $Z'$ boson with a mass of
$2\TeV$ and of negligible width (though not necessarily a likely
scenario, it is adequate for examining the kinematical aspects of
interest here). In fig.~\ref{fig:Zpmass} we show the mass distribution
as obtained directly at hadron level and also after the subtraction
procedure, eq.~(\ref{eq:pt-correct}). In the case of events without
pileup, the subtraction removes just the moderate underlying event
contribution, with limited effect on the sharpness of the peak.
With pileup, the mass distribution is shifted and broadened
significantly. The subtraction brings the peak mass back into accord
with the value measured without pileup, and restores a significant
part of the resolution that had been lost. This is quantified in
table~\ref{tab:Zp}, which includes results also for the
Cambridge/Aachen and SISCone ($f=0.5$) algorithms, all for $R=0.7$,
and shows the effectiveness of the subtraction there too.

The last of our high-luminosity LHC $pp$ pileup studies concerns top quark
reconstruction.  We simulate a sample of $t\bar t$ events that decay
to $\ell^+\nu_\ell b + q \bar q' \bar b$, and assume that both the $b$
and $\bar b$ jets have been tagged. We then look for the two hardest
of the non-tagged jets and assume they come from the $W \to q\bar q'$
decay. For simplicity we eliminate the combinatorial background in the
top reconstruction by pairing the hadronic $W$ with the $b$ or $\bar
b$ according to the lepton charge.\footnote{While this may not be
  realistic experimentally, it should be largely irrelevant to the
  question of how pileup and subtraction affect the kinematics of the
  reconstruction.} %
The resulting invariant mass distributions for the $W$ and top are
shown in fig.~\ref{fig:Wttbar}, for events with and without
(high-luminosity) pileup. As in fig.~\ref{fig:Zpmass} we show them as
measured directly at hadron level and also after the subtraction
procedure, eq.~(\ref{eq:pt-correct}). Despite the small $R$ value, $0.4$, the
pileup still significantly smears and shifts the peak. The subtraction
once again allows one to recover the original distributions to a large
extent, this independently of the choice of jet algorithm.

\begin{figure}[t]
  \includegraphics[width=0.48\textwidth]{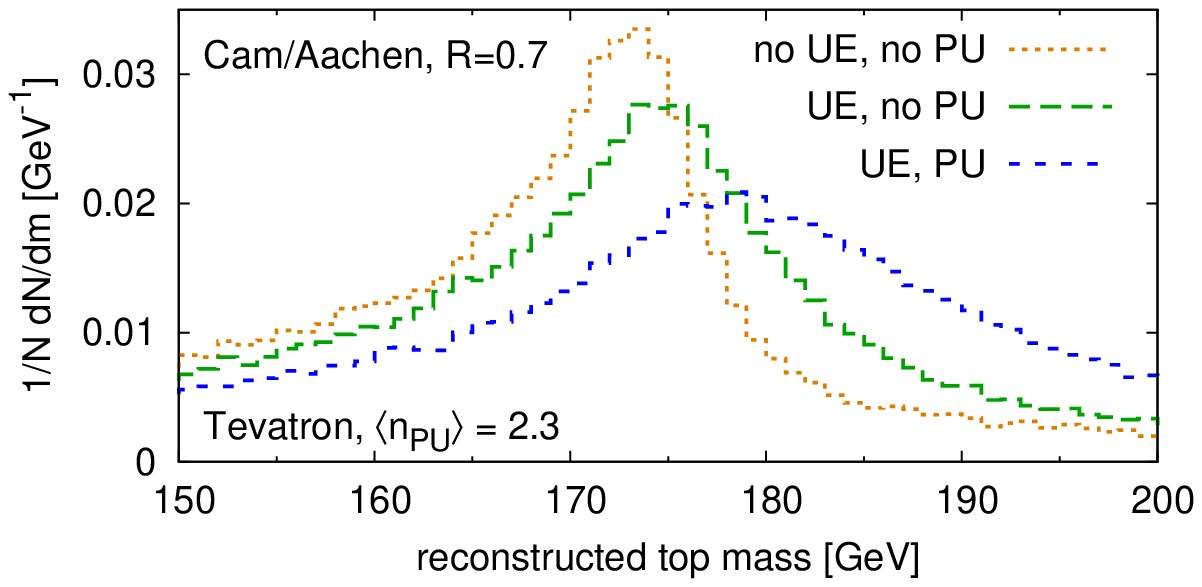}\hfill
  \includegraphics[width=0.48\textwidth]{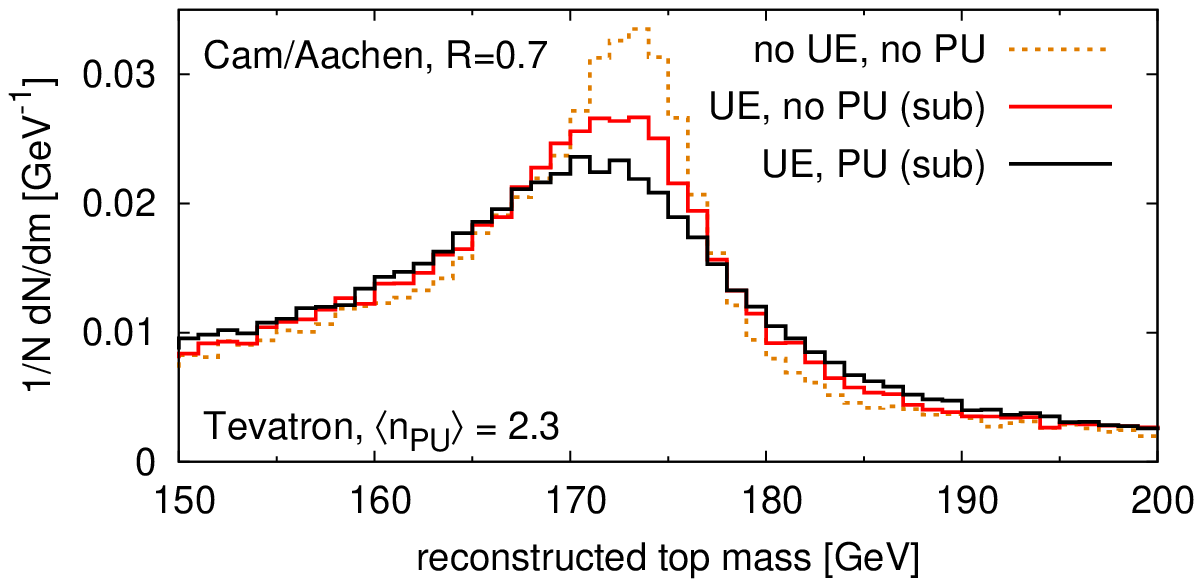}
  \caption{\small Subtraction in the context of top reconstruction for
    Tevatron kinematics (simulated with Pythia~6.325, $m_t=175\GeV$), in the lepton +
    jets decay channel ($\rho$ extracted with the $k_t$ algorithm and
    $R=0.5$, using jets with $|y|<3$). }
  \label{fig:Tevatron}
\end{figure}

For our final two examples we consider situations other than
high-luminosity $pp$ collisions at LHC. One will have relatively low
background contamination, the other very high contamination. As we
shall see, they can both be considered as particularly challenging,
albeit for opposing reasons.

Firstly we examine $t\bar t$ production at the Tevatron with a modest
pileup contribution $\langle n_{\text{PU}} \rangle \simeq 2.3$, which
corresponds roughly to current instantaneous
luminosities~\cite{CDFkt}. This is challenging for two reasons: while
the mean value of the background contamination is smaller than at the LHC, its
relative fluctuations ($\sigma/\rho$) are considerably larger; also,
the neglected loss term of eq.~(\ref{eq:deltapt}) is suppressed
relative to the main subtraction by $\as \ln p_t/(A\rho)$, which is no longer
as small a parameter as it was for large $\rho$.
To complicate the problem a little more, we enhance the effect of the
UE and PU by using a relatively large value of the jet radius, $R=0.7$.
Figure~\ref{fig:Tevatron} shows the unsubtracted reconstructed mass
distributions in the left-hand panel, as simulated with Pythia, and
the subtracted results in the right-hand panel, using the
Cambridge/Aachen algorithm. For reference we include also the
distribution obtained with neither pileup nor underlying event. One
sees that the subtraction brings one rather close to this result. The
same feature is observed for the SISCone algorithm, while for the
$k_t$ algorithm the coincidence is not as quite good, there being an
over-subtraction of a couple of GeV on the position of the peak,
perhaps attributable to a slightly greater fragility of
$k_t$-algorithm jets (\ie a larger typical contribution from the third
term of eq.~(\ref{eq:deltapt})~\cite{css-area}).
We have also examined top reconstruction for low-luminosity LHC
running. We find results that are rather similar to those for
the Tevatron.

\begin{figure}[t]
    \includegraphics[width=0.48\textwidth,height=0.31\textwidth]{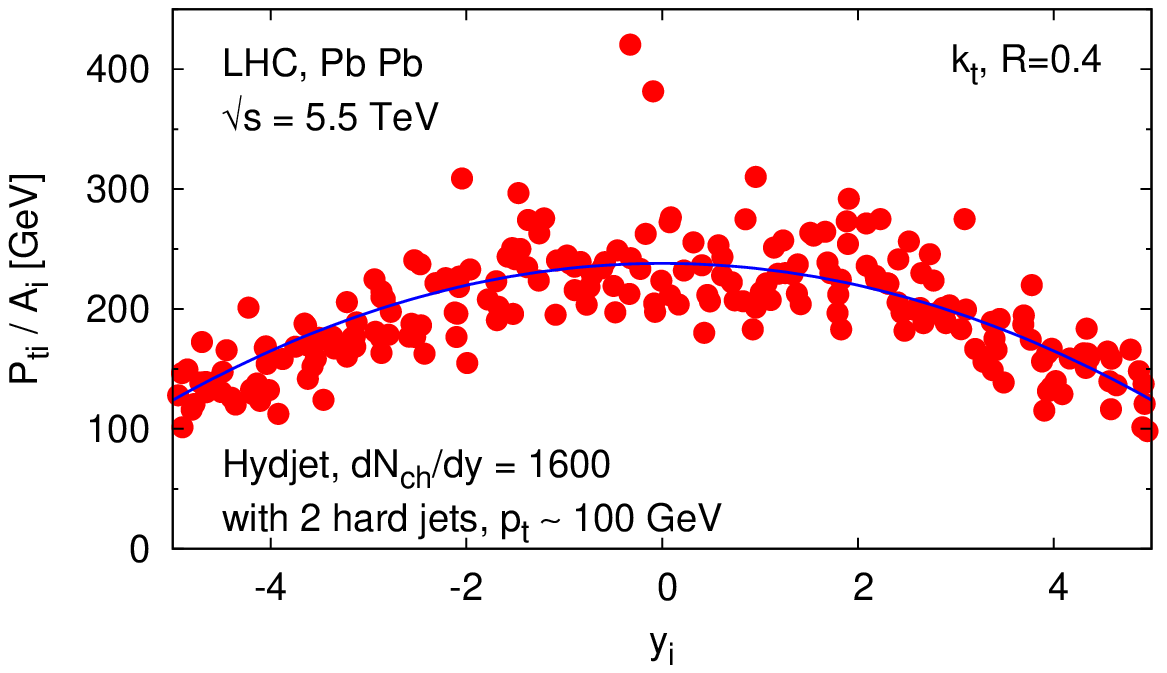}\hfill
    \includegraphics[width=0.48\textwidth,height=0.31\textwidth]{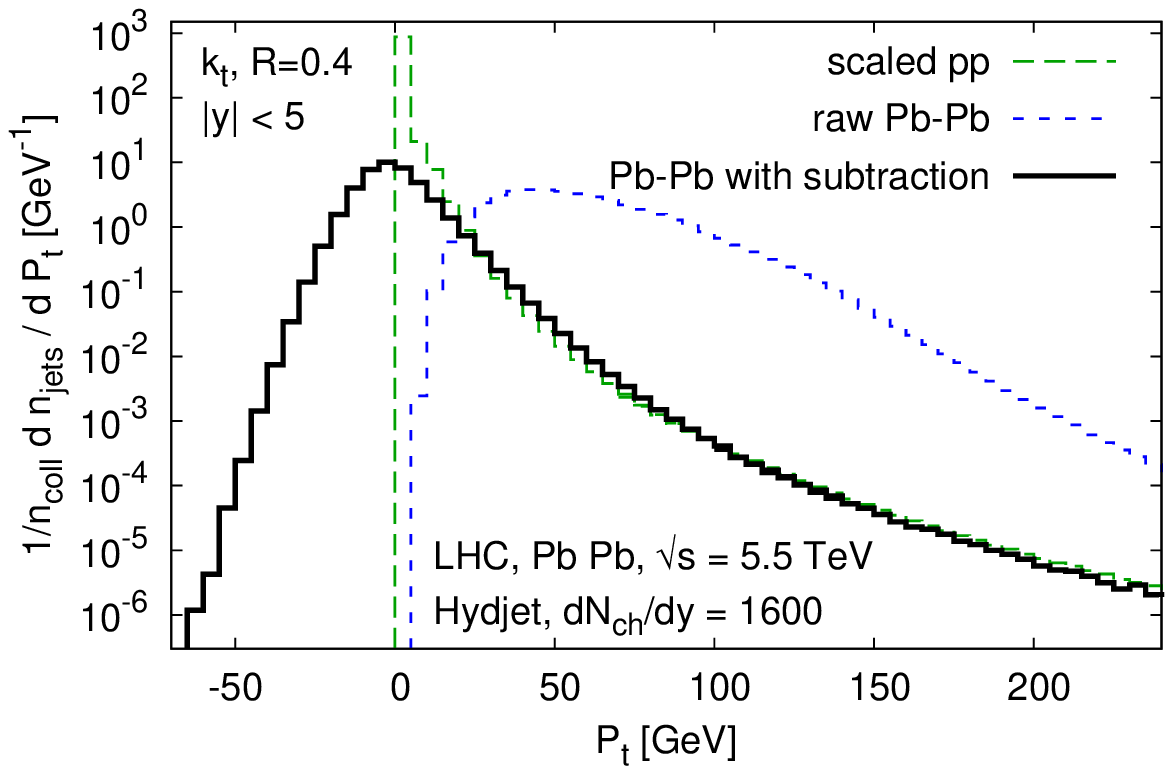}
    \caption{\small Left: scatter plot of $p_{ti}/A_i$ for jets in a
      central Pb\,Pb collision at the LHC (simulated with Hydjet~1.1,
      without quenching),
      together with a parabolic fit.  Right: inclusive jet spectrum,
      scaled from $pp$ collisions, as would be measured directly in
      Pb\,Pb and after subtraction.\vspace{-0.5cm}}
  \label{fig:HI}
\end{figure}

The high background contamination example that we consider for our
procedure is that of heavy-ion collisions at the LHC, where a single
\pbpb collision produces a diffuse background with a transverse
momentum density $\rho$ about ten times larger than that of high
luminosity $pp$ pileup, \ie $\sim 250$~GeV.  This is challenging
because jets normally considered as hard, with a $p_t$ of order
50--100~GeV or more, can be swamped by the background.
Fig.~\ref{fig:HI} (left), the analogue of
fig.~\ref{fig:pt_over_area-v-rap}a but now for a central \pbpb
collision simulated with Hydjet~\cite{hydjet}, shows that most jets
still lie in a collimated band. This band however depends noticeably
on rapidity $y$ (for reasons related to the heavy-ion collision
dynamics), so rather than using a constant $\rho$, we introduce a
function $\rho_\text{HI}(y)$. We parametrise it as $\rho_\text{HI}(y)
= a + b y^2$, where the coefficients $a$ and $b$ are to be fitted for
each event.\footnote{Note that: 1) the systematically large $p_t$ of
  the many background jets means that the fit will only be minimally
  biased by any truly hard jets --- therefore it is not necessary to
  resort to techniques such as the use of the median in order to
  obtain robust results for $\rho$; 2) for non-central collisions
  $\rho$ is expected to have non-negligible dependence also on $\phi$,
  and one may generalise both fit and median-based procedures to deal
  with these more complex situations.} %
Despite the huge background, our subtraction procedure remains
effective even at moderate $p_t$'s, as illustrated by the inclusive
jet spectrum shown in fig.~\ref{fig:HI} (right). One notes also the
presence of a steep tail at negative $p_t$. It has the same origin as
the negative shifts in fig.~\ref{fig:sub-correl-2tile}b--d, \ie
principally the fact that local fluctuations in the background level
cause some jets' contamination to be lower than $A \rho$. The width of
this tail at negative $p_t$ (note the logarithmic ordinate scale)
provides an alternative estimate on the resolution associated with the
subtraction.

\section{Conclusions}
\label{sec:concl}

We have here introduced a new procedure for correcting jets
for pileup and underlying event contamination.  It is based on the use
of infrared-safe jet algorithms and the novel concept of jet area. On
an event-by-event basis it estimates the level of the diffuse
background in the event and then uses this estimate to correct each
jet in proportion to its area.
The procedure is entirely data driven, essentially parameter-free, it
does not necessitate Monte Carlo corrections in order to give the
correct results and it provides an associated estimate of the
uncertainty on the subtraction.
A full validation of the method would require that one carry out tests
in an experimentally realistic context, accounting in particular for
all detector-related effects and limitations. This is beyond the scope
of this article, where we have restricted our attention to
hadron-level investigations.
Tests have been performed on simulated events in a range of cases:
moderate luminosity $p\bar p$ collisions at the Tevatron, low and
high-luminosity $pp$ collisions at the LHC and \pbpb collisions at the
LHC.
Despite its relative simplicity our approach is successful in all the
cases we have examined. This is true both for simple quantities like
individual jet transverse momenta, and for more complex analyses, \eg
top mass reconstruction, where we recover the correct momentum and
mass scales, and significantly improve the resolution compared to the
uncorrected results.

\section*{Acknowledgements}

We thank G\"unther Dissertori and Monica Vazquez Acosta 
for helpful suggestions and comments 
and David d'Enterria, Peter Jacobs and Christof Roland for useful
discussions in the context of heavy-ion collisions.
We also thank Gregory Soyez for his contribution to the development of
several of the area-related tools in FastJet.
Finally, we thank the referee for many insightful comments and
suggestions.
This work has been supported in part by grant ANR-05-JCJC-0046-01 from
the French Agence Nationale de la Recherche.

\end{document}